\documentclass[physrev,twocolumn,floatfix,superscriptaddress,longbibliography]{revtex4-2} 
\usepackage[utf8]{inputenc}
\usepackage{graphicx}
\usepackage{bigstrut}
\usepackage{amsmath}
\usepackage{xcolor}
\usepackage{multirow}
\usepackage{hyperref}
\hypersetup{colorlinks=true,allcolors=blue}
\usepackage{orcidlink}
\definecolor{light-gray}{gray}{0.98}
\newcommand{\code}[1]{\colorbox{light-gray}{\texttt{#1}}}
\usepackage{xr}
\definecolor{color1}{rgb}{0.858, 0.188, 0.478}

\begin{document}
\title{A critical examination of robustness and generalizability of machine learning prediction of materials properties}

\author{Kangming Li\orcidlink{0000-0003-4471-8527}}
\email[Correspondence: ]{kangming.li@utoronto.ca}
\affiliation{Department of Materials Science and Engineering, University of Toronto, 27 King’s College Cir, Toronto, ON, Canada.}

\author{Brian DeCost\orcidlink{0000-0002-3459-5888}}
\affiliation{Material Measurement Laboratory, National Institute of Standards and Technology, 100 Bureau Dr, Gaithersburg, MD, USA.}

\author{Kamal Choudhary\orcidlink{0000-0001-9737-8074}}
\affiliation{Material Measurement Laboratory, National Institute of Standards and Technology, 100 Bureau Dr, Gaithersburg, MD, USA.}
\affiliation{Theiss Research, La Jolla, CA 92037, USA.}

\author{Michael Greenwood}
\affiliation{Canmet MATERIALS, Natural Resources Canada, 183 Longwood Road south, Hamilton, ON, Canada.}

\author{Jason Hattrick-Simpers\orcidlink{0000-0003-2937-3188}}
\affiliation{Department of Materials Science and Engineering, University of Toronto, 27 King’s College Cir, Toronto, ON, Canada.}

\begin{abstract}
Recent advances in machine learning (ML) methods have led to substantial improvement in materials property prediction against community benchmarks, but an excellent benchmark score may not imply good generalization performance. Here we show that ML models trained on the Materials Project 2018 (MP18) dataset can have severely degraded prediction performance on new compounds in the Materials Project 2021 (MP21) dataset. We demonstrate performance degradation in graph neural networks and traditional descriptor-based ML models for both quantitative and qualitative predictions. We find a plausible source of the predictive degradation is due to the distribution shift between the MP18 and MP21 versions. This is revealed by the uniform manifold approximation and projection (UMAP) of the feature space. We then show that the performance degradation issue can be foreseen using a few simple tools. Firstly, the UMAP can be used to investigate the connectivity and relative proximity of the training and test data within feature space. Secondly, the disagreement between multiple ML models on the test data can illuminate out-of-distribution samples. We demonstrate that the simple yet efficient UMAP-guided and query-by-committee acquisition strategies can greatly improve prediction accuracy through adding only 1~\% of the test data. We believe this work provides valuable insights for building materials databases and ML models that enable better prediction robustness and generalizability.
\end{abstract}

\maketitle

\section{Introduction}

The use of machine learning (ML) has been increasingly popular in the materials science community~\cite{Butler2018,Vasudevan2019,Morgan2020a,decost2020scientific,Hart2021,Stach2021,Choudhary2022,schleder2019dft,green2022autonomous,kalinin2022machine,krenn2022scientific}. 
Central to the training of machine learning models is the need for findable, accessible, interoperable, and reusable (F.A.I.R.)~\cite{wilkinson2016fair} materials science datasets. High-throughput density functional theory (DFT) calculations have proven to be an efficient and reliable way to generate materials property data, screen the target materials space and accelerate materials discovery~\cite{jain2011high,Saal2013,garrity2021database,horton2019high,armiento2011screening}. 
Concentrated community efforts have led to the curation of large DFT databases for various materials properties, e.g., Materials Project~\cite{Jain2013}, Automatic FLOW for Materials Discovery~\cite{Curtarolo2012}, Open Quantum Materials Database~\cite{Saal2013}, JARVIS-DFT~\cite{Choudhary2020}. The availability of large materials databases has fueled the development and application of machine learning methods based on chemical formula or atomic structures, including traditional ML models with preselected features sets~\cite{bartok2013representing,de2016statistical,ouyang2018sisso,Schutt2014,Faber2015,Ward2016,Ward2017a,Ward2018,choudhary2018machine} and neural networks with automatic feature extraction~\cite{
jha2018elemnet,Xie2018,Chen2019,DeBreuck2021,Choudhary2021,schmidt2021crystal,Ihalage2022,Dunn2020,chen2021atomsets,choudhary2022unified,chen2022universal}.

The continuously improved performance of ML models in the DFT database benchmarks shows the great potential of using these models as the surrogate of computationally expensive DFT calculations to explore unknown materials~\cite{Dunn2020}. However, there are reasons to remain cautious, particularly for the generalization performance of the trained ML models~\cite{stein2022advancing}. First, the current DFT databases still cover only a very limited region of the potential materials space~\cite{Kirkpatrick2004,Davies2016}. Some databases may be the results of mission-driven calculations and therefore be more focused on certain types of materials or structural archetypes, leading to biased distributions~\cite{Griffiths2021,Breuck2021,Kumagai2022}. In addition, data distributions may shift even between different versions of an actively expanding database, due to a change of their focus with time. 
While the common practice is to train and validate ML models on the latest databases, we are unaware of any systematic study examining whether these models can predict reasonably (or at least qualitatively) well the properties of new materials added in the future database versions. Such an examination is critical for assessing the ``maturity" of a database (whether it is sufficiently representative of the materials space) and the robustness of the resulting ML prediction, both of which are essential for building the trust in the use of these ML models. 

Since the current databases may not yet offer an unbiased and sufficiently rich representation of the potential materials space, the performance scores of an ML model evaluated from a random train-validation-test split may be an optimistic estimate of the true generalization performance~\cite{kauwe2020can,xiong2020evaluating,zahrt2020cautionary}. While the latter may be estimated more properly from grouped cross-validations (CV)~\cite{Ren2018,Meredig2018,zhao2022limitations}, finding a well-defined method for grouping data is not always trivial and depends on the pre-selected input features, which may not be the optimal way to find the most physically relevant grouping~\cite{Dunn2020}. On the other hand, one may consider it safer to limit the use of an ML model to its applicability domain, or the interpolation region~\cite{Kumagai2022}. However, in a high-dimensional compositional-structural feature space as is encountered in materials science, it is challenging to properly define an interpolation region and to determine when the model is extrapolating. 

In this work, we highlight the limitations of the current ML methods in materials science for predicting out-of-distribution samples, by showing that ML models pretrained on the Materials Project~\cite{Jain2013} 2018 database have unexpectedly acute performance degradation on the latest database. Such performance degradation can occur in the deployment stage of any ML model and degrades community trust in their validity. Therefore, we also provide solutions for diagnosing, foreseeing and addressing the issue, and discuss ways to improve prediction robustness and generalizability.

The paper is organized as follows. First, we examine the performance of a state-of-the-art neural network, with a comparison to traditional ML models. Next, we analyze the observed performance degradation in terms of the dataset's feature space. We then discuss different methods based on the dataset's representation and model predictions to foresee the generalization issue. Finally, we propose ways to improve prediction robustness for materials exploration. 

\section{Results and discussion}

\subsection{Failure to generalize in new regions of materials space}
\label{sec:raise_issue}

Formation energy ($E_f$) is a fundamental property that dictates the phase stability of a material. Formation energy prediction is a basic task for ML models used in materials science, including traditional descriptor-based models~\cite{Ward2016,Ward2017a,Ward2018,Bartel2020} and neural networks~\cite{Xie2018,Chen2019,DeBreuck2021,Choudhary2021,Ihalage2022}. Among them, graph neural network (GNN) models with atomistic structures as inputs are currently considered to have the state-of-the-art performance~\cite{Choudhary2022}. Here we consider the Atomistic LIne Graph Neural Network (ALIGNN), an architecture that performs message passing on both the interatomic bond graph and its line graph corresponding to bond angles~\cite{Choudhary2021}. The ALIGNN model shows the best performance in predicting the Materials Project~\cite{Jain2013} formation energy according to the Matbench~\cite{Dunn2020} leader-board, we therefore choose it as the representative GNN model for the subsequent performance evaluation.

We use the ALIGNN model pretrained on the Materials Project 2018.06.01 version (denoted as MP18), which contains 69239 materials and has been used for benchmarking GNN models in the recent papers~\cite{Chen2019,Choudhary2021,DeBreuck2021}. In the original ALIGNN paper, a 60000–5000–4239 train-validation-test split of the MP18 dataset was used, achieving a mean absolute error (MAE) of 0.022 eV/atom for the test set~\cite{Choudhary2021}.

We use the MP18-pretrained ALIGNN model (ALIGNN-MP18) to predict the formation energies of the new structures in the latest (2021.11.10 version) Materials Project database (denoted as MP21). Instead of testing on the whole MP21 dataset, we consider the scenario where we want to apply ML models to explore a particular material subspace of interest. In this work, we define the alloys of interest (AoI) as the space formed by the first 34 metallic elements (from Li to Ba) and the alloys formed exclusively by these elements. In the MP21 dataset, there are 7800 AoI, 2261 (or 29~\%) of which already appear in the MP18 dataset while the rest are not contained within MP18. Therefore, we consider those 2261 alloys as the AoI in the training set, and the rest that appear only in the MP21 as the AoI in the test set. 

A description of the MP18 dataset, and the AoI data is given in Table~\ref{tab:dataset_descrip}. We note that 
the mean absolute deviation (MAD) and the standard deviation (STD) of the data correspond to the mean absolute error (MAE) and the root mean square error (RMSE) of a baseline model whose prediction for every structure is equal to the mean of the training data.
\begin{table}
  \centering
  \caption{Description of the MP18 data, and the AoI data in the MP18 and MP21 datasets. The number of entries, the minimum, maximum, mean absolute deviation and standard deviation of formation energies (in eV/atom) are given.
  }
  \label{tab:dataset_descrip}
\begin{ruledtabular}
\begin{tabular}{lrrrrr}
         & Entries & Min    & Max   & MAD   & STD  \bigstrut[b] \\
\hline
MP18     & 69239   & -4.522 & 4.389 & 0.926 & 1.072 \\
MP18 AoI & 2261    & -1.090 & 1.575 & 0.230 & 0.313 \\
MP21 AoI & 7800    & -1.090 & 4.416 & 0.440 & 0.751
\end{tabular}
\end{ruledtabular}
\end{table}

\begin{figure}
\centering
	\includegraphics[width=0.7\linewidth]{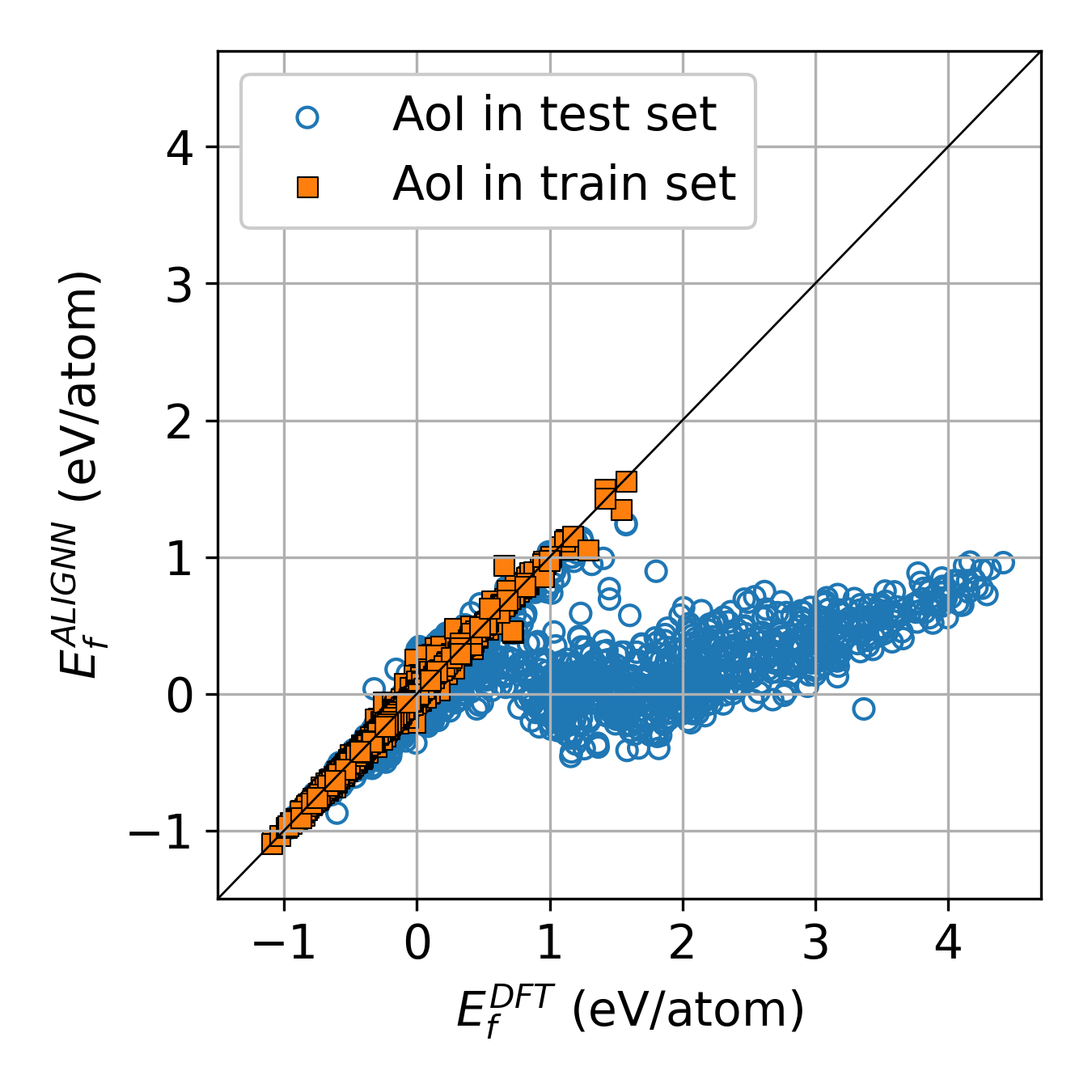}
       \includegraphics[width=0.7\linewidth]{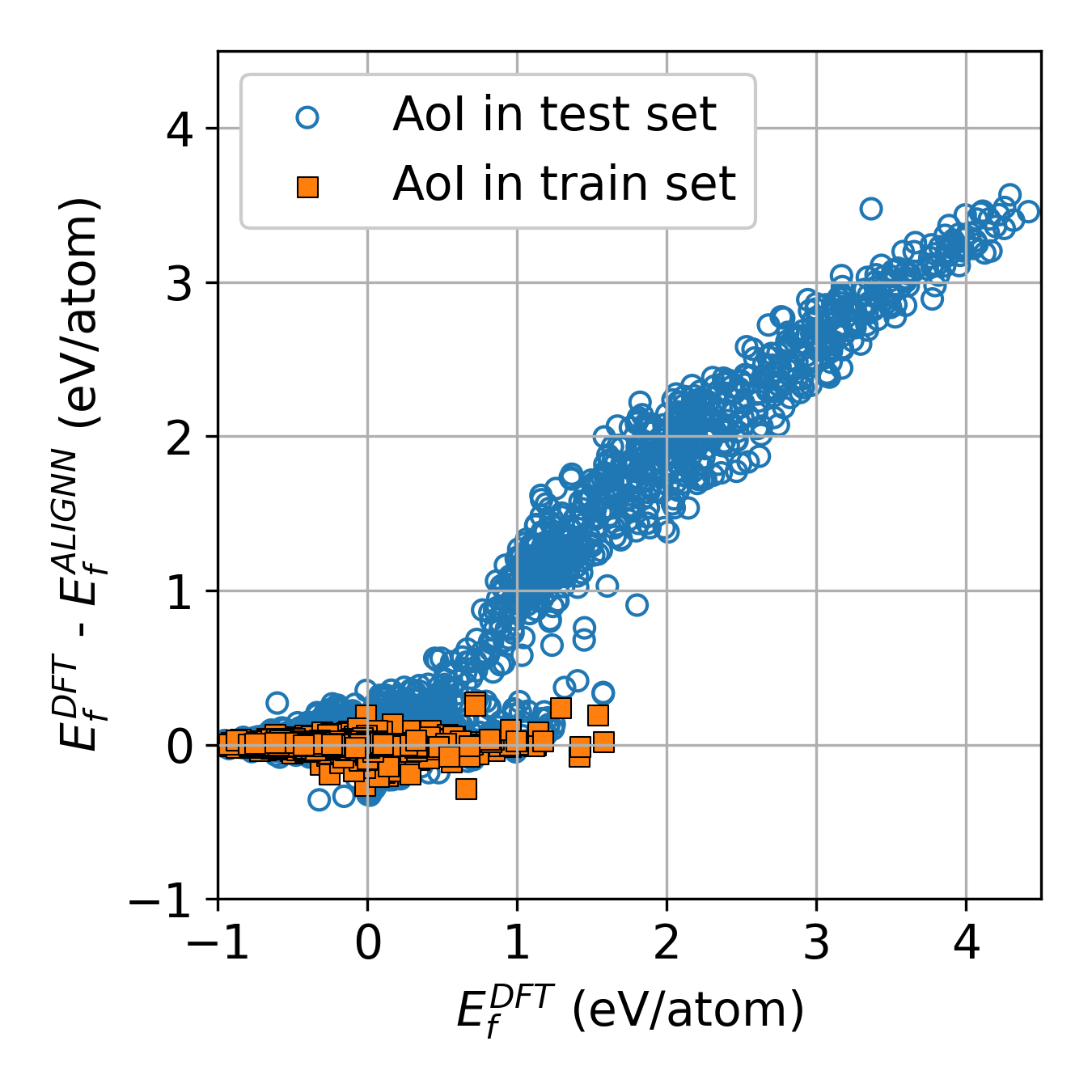}
	\caption{\label{fig:raise_issue_alignn} Parity plot and prediction errors of the ALIGNN-MP18 model. The diagonal line $Y=X$ in the upper panel represents the perfect agreement.}
\end{figure}

Fig.~\ref{fig:raise_issue_alignn} shows the ALIGNN-MP18 performance on the formation energy predictions of the AoI. For the AoI in the training set, the ALIGNN-MP18 predictions agree well with the DFT values, with a MAE of 0.013 eV/atom. For the AoI test data, while there is still a reasonable agreement for the structures with $E_f^{\text{DFT}}$ below 0.5 eV/atom, the ALIGNN-MP18 model strongly underestimates the formation energies for a significant portion of the structures whose $E_f^{\text{DFT}}$ are above 0.5 eV/atom. In the latter case, the prediction errors range from 0.5 eV/atom to up to 3.5 eV/atom, which is 23 to 160 times larger than the MP18 test MAE of 0.022 eV/atom. Indeed, the prediction errors are nearly as large as $E_f^{\text{DFT}}$ for those alloys, indicating that the ALIGNN-MP18 predictions fail to even qualitatively match the DFT formation energies. For reference, the MAE and the coefficient of determination ($R^2$ score) for the AoI test set are 0.297 eV/atom and 0.194, respectively (Table~\ref{tab:models_metrics}).

It can be seen from Fig.~\ref{fig:raise_issue_alignn} that the ALIGNN-MP18 predictions are largely restricted to the value range below 1 eV/atom. Indeed, despite a large formation energy range (from -4.3 eV/atom to 4.4 eV/atom) of the whole MP18 dataset, most of the formation energies of the AoI in the MP18 lie between -1 eV/atom to 1 eV/atom. Therefore, it is not surprising that the ALIGNN-MP18 predictions are limited by the range of the formation energies of the AoI training set. However, it is unexpected to observe that the strong underestimation by the ALIGNN-MP18 model already occurs in the the formation energy range of 0.5 eV/atom to 1 eV/atom. For alloys with formation energies above 1 eV/atom, the ALIGNN-MP18 model predicts values that are well below the upper bound of formation energies in the training set, some of which are even negative. Consequently, the test set performance issue of the ALIGNN-MP18 model cannot be explained by the bounded energy range of the AoI in the training set. The origin of the issue will be discussed in the next section.


To verify whether the performance issue is common to other ML models, we perform the same training and test procedures with traditional descriptor-based ML models. To do so, we first use Matminer~\cite{Ward2016,Ward2017a,Ward2018} to extract 273 features based on compositions and structures for the whole MP18 dataset and the alloys in MP21. Then, we down select features by sequentially dropping highly correlated features using a Pearson's R of 0.7 as the threshold, reducing the final number of features to 90. These 90 features are used for subsequent traditional ML model training and other analysis throughout this work.

Here we consider three traditional regression models: the gradient-boosted trees as implemented in \code{XGBoost} (XGB)~\cite{xgboost}, random forests (RF) as implemented in \code{scikit-learn}~\cite{sklearn}, and linear forests (LF) as implemented in \code{linear-forest}~\cite{linearforest,Zhang2017}. XGB builds sequentially a number of decision trees in a way such that each subsequent tree tries to reduce the residuals of the previous one. RF is an ensemble learning technique that combines multiple independently built decision trees to improve accuracy and minimize variance. LF combines the strengths of the linear and RF models, by firstly fitting a linear model (in this work a Ridge model) and then building a RF on the residuals of the linear model. 

The motivation of using the traditional models for understanding the ALIGNN-MP18 performance issue is three-fold. 
First, do traditional ML models fail to generalize as well, or is this failure unique to neural networks? 
Second, traditional models can provide more interpretability than neural networks and can be used as surrogate models of the ALIGNN in the subsequent analysis. Finally, traditional models are computationally much easier to train than large neural networks, allowing us to perform more detailed statistical examinations. In fact, the reference implementation of ALIGNN-MP18~\cite{Choudhary2021} required a total compute cost of 28 GPU hours plus 224 CPU hours for training on the MP18 dataset. For comparison, the same training with traditional ML models takes 0.02 CPU hours (4 orders of magnitude less compute than the ALIGNN). 

\begin{table*}[htbp]
  \centering
  \caption{Comparison of MAE (in eV/atom), RMSE (in eV/atom) and coefficient of determination ($R^2$) between different ML models. The metrics for the MP18 dataset are obtained for the test set following the same 60000–5000–4239 train-validation-test split as in the ALIGNN-MP18 paper~\cite{Choudhary2021}. The metrics for the new alloys in the MP21 are obtained with the predictions of the MP18-pretrained models. The last two columns show the ratio of the prediction error of the new MP21 alloys compared to that of the MP18.
  }
  \label{tab:models_metrics}
\begin{ruledtabular}
\begin{tabular}{lcccccccccc}
      & \multicolumn{3}{c}{MP18} &       & \multicolumn{3}{c}{New AoI in MP21} &       & \multicolumn{2}{c}{Ratio of metrics} \bigstrut[b]\\
\cline{2-4}\cline{6-8}\cline{10-11}      & MAE   & RMSE  & $R^2$    &       & MAE   & RMSE  & $R^2$    &       & \multicolumn{1}{c}{MAE} & \multicolumn{1}{c}{RMSE} \bigstrut\\
\cline{1-4}\cline{6-8}\cline{10-11}ALIGNN-MP18 & 0.022 & 0.052 & 0.999 &       & 0.297 & 0.747 & 0.194 &       & 13.5  & 14.4 \bigstrut[t]\\
XGB   & 0.075 & 0.137 & 0.984 &       & 0.239 & 0.537 & 0.582 &       & 3.2   & 3.9 \\
RF    & 0.088 & 0.165 & 0.977 &       & 0.382 & 0.879 & -0.119 &       & 4.3   & 5.3 \\
LF  & 0.108 & 0.179 & 0.972 &       & 0.327 & 0.606 & 0.469 &       & 3.0   & 3.4 \\
\end{tabular}%
\end{ruledtabular}
\end{table*}

\begin{figure*}[htbp]
    \includegraphics[width=\linewidth]{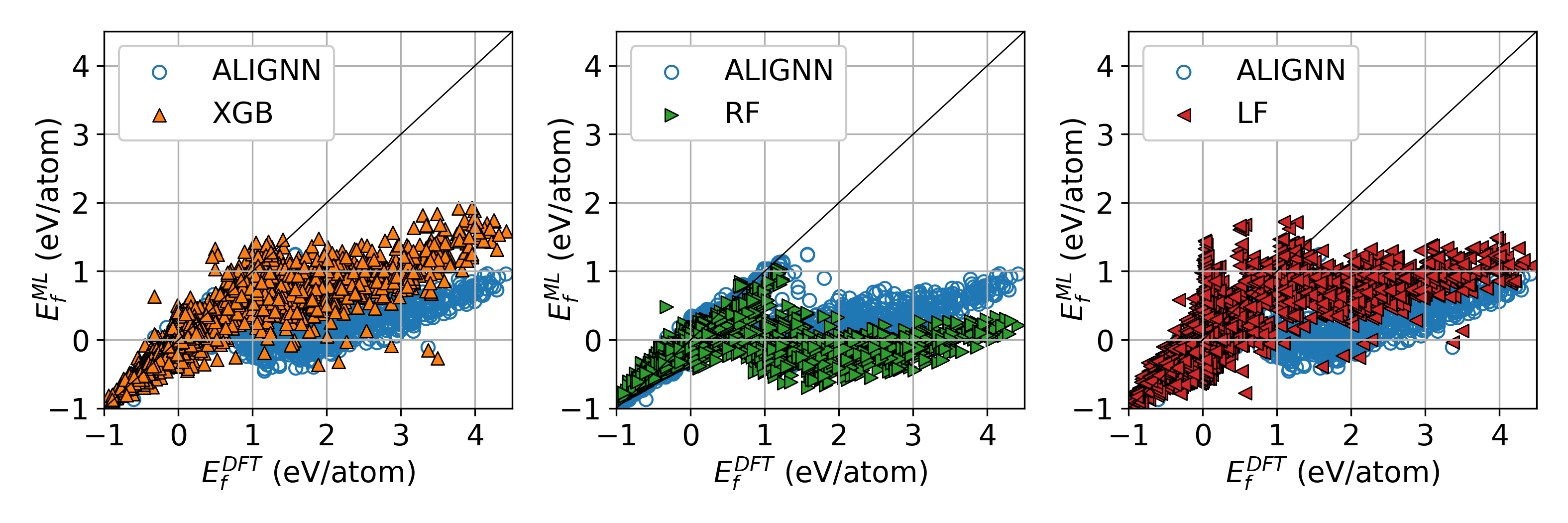}
    \caption{\label{fig:raise_issue_compare} Comparison of MP18-pretrained model performance on the AoI test set.}
\end{figure*}

First, the XGB, RF and LF models are hypertuned, trained and tested with the same train-validation-test split of the MP18 as for the ALIGNN model. Then, the models are trained on the MP18 and tested on the new AoI in the MP21. Comparisons of their performance metrics and predictions are shown in Table~\ref{tab:models_metrics} and Fig.~\ref{fig:raise_issue_compare}, respectively. 

Table~\ref{tab:models_metrics} shows that the MP18 test set MAE of the traditional models are three to five times larger than that of the ALIGNN-MP18. This is consistent with literature findings that neural networks usually outperform traditional models in various benchmarks of large materials databases~\cite{Dunn2020,Choudhary2022}. However, evaluating model performance from the random train-validation-test split is based on the assumption that data distributions are identical for the training and test sets, which may not hold when exploring new materials. Therefore, such performance scores are not good estimates of the model true generalizability~\cite{Dunn2020,Meredig2018}. Indeed, when the models are applied to the new AoI in MP21, the large performance difference between traditional and ALIGNN models disappear. More strikingly, XGB outperforms ALIGNN in terms of the MAE, RMSE and $R^2$ scores, whereas LF outperforms ALIGNN in terms of the RMSE and $R^2$ scores. An equal footing  comparison of the extrapolation performance should also take into account the complexity and capacity of the models. A more consistent comparison may be to compute the ratio of the performance metrics obtained with the training set and the test set, which are shown as the last two columns in Table~\ref{tab:models_metrics}. The performance degradation of the traditional models is less severe than that of the ALIGNN model.  

Fig.~\ref{fig:raise_issue_compare} gives a more detailed comparison of the prediction performance on the MP21 new alloys. Compared to the ALIGNN model, the XGB model leads to larger errors in the $E_f^{\text{DFT}}$ range below 0.5 eV/atom, but performs considerably better for predicting high-energy alloys, of which there are fewer structures that are misclassified as having negative formation energies. On the other hand, the RF model performs similarly to the ALIGNN model in the $E_f^{\text{DFT}}$ range below 0.5 eV/atom but worse than the latter for high-energy structures. Interestingly, the LF model, in which the linear model is first fitted before training the RF model, improves the predictions for high-energy structures to an extent similar to the XGB model. The better RMSE scores for the XGB and LF models are attributed to the less degraded predictions for those high-energy structures. 

The above discussion of Table~\ref{tab:models_metrics} and Fig.~\ref{fig:raise_issue_compare} shows that the performance degradation issue observed in the ALIGNN model also occurs in other traditional descriptor-based models, but the performance degradation can be quite different, with the XGB and LF models demonstrating less performance degradation. In the following sections, we will reveal the origin of the performance degradation and the reasons behind the better generalizability of the XGB and LF models.

\subsection{Diagnosing generalization performance degradation}
\label{sec:explain_issue}
In the previous section, we have shown that the performance issue on the AoI test set is common to different ML models, indicating that it is likely related to the distribution shift between the training and the test sets. For instance, the test set may cover compositions or structures that lie far away from the training set. Here we show how to diagnose this issue in a holistic and detailed manner, and discuss some important insights resulting from this analysis.

\begin{figure*}[!hbtp]
	\includegraphics[width=\linewidth]{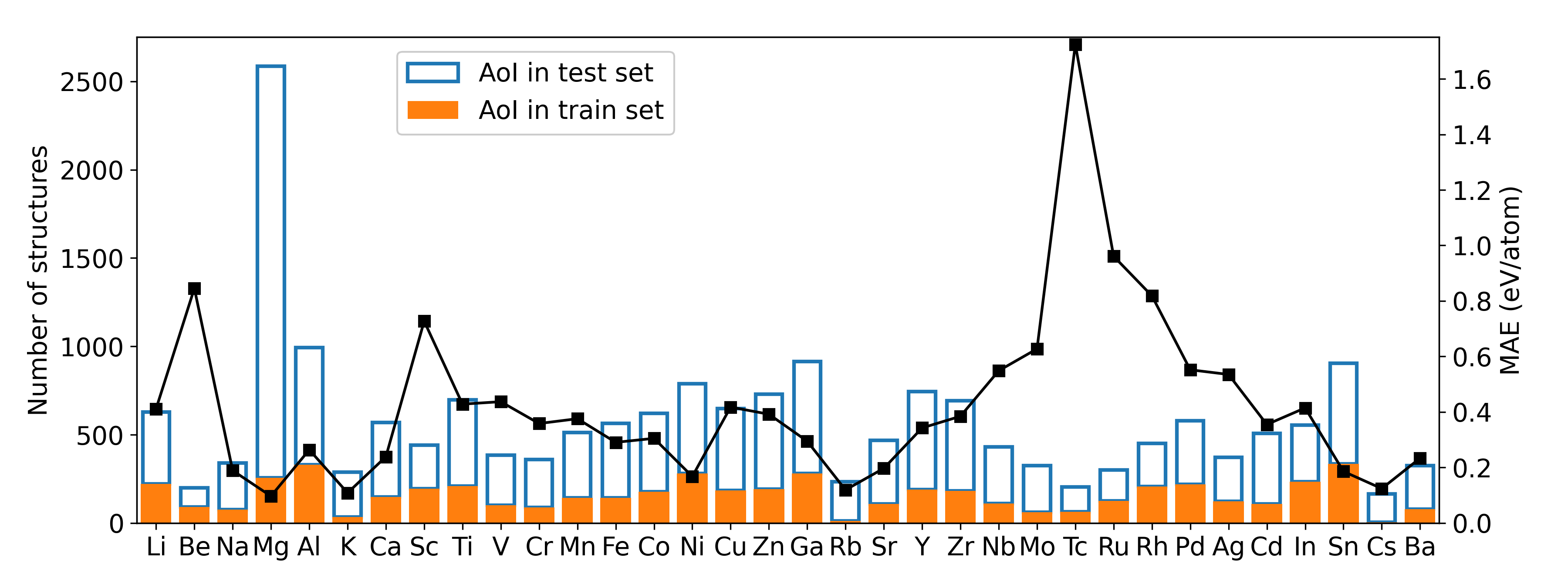}	\caption{\label{fig:explain_issue_elements_dist} Number of AoI containing a given element. The line plot (with respect to the right Y axis) indicates the MAE by the ALIGNN-MP18 on the AoI test set.}
\end{figure*}

We start by comparing the distributions of some basic compositional and structural features between the MP18 and MP21 datasets. In Fig.~\ref{fig:explain_issue_elements_dist} we count for each element $X$ of 34 metallic elements the number of $X$-containing AoI in the training and the test set. We also plot the MAE of ALIGNN-MP18 for the corresponding $X$-containing AoI in the test set to investigate potential correlations between large MAE and elements that are underrepresented in the AoI training set. We find that although there are few AoI that contain elements such as K, Rb and Cs in the training set, the corresponding test MAE are actually rather small. Indeed, we find a Spearman's rank correlation coefficient ($r_S$) of 0.06, i.e., negligible correlation, between the test MAE of $X$-containing AoI and the number of $X$-containing AoI in the training set. Meanwhile, we find a weak anti-correlation ($r_S$ equal to -0.42) between the test MAE of $X$-containing AoI and the number of \textit{all} $X$-containing structures (i.e., AoI and Non-AoI) in the training set, although such a correlation vanishes above a threshold of 1000 $X$-containing structures (Supplemental Information Fig.~\ref{S-fig:explain_issue_nelements}). This suggests that chemically less relevant data can still inform ML models and may reduce generalization errors in a target subspace, though to a limited extent.

Another basic composition-related feature is the number of elements contained in a structure. We find that the majority of the AoI in the training and the test sets are binary and ternary systems. The poorly predicted structures are the ternary alloys and some binary ones that have a $E_f^{\text{DFT}}$ larger than 0.5 eV/atom (Supplemental Information Fig.~\ref{S-fig:explain_issue_nelements}). 

\begin{figure}
	\includegraphics[width=\linewidth]{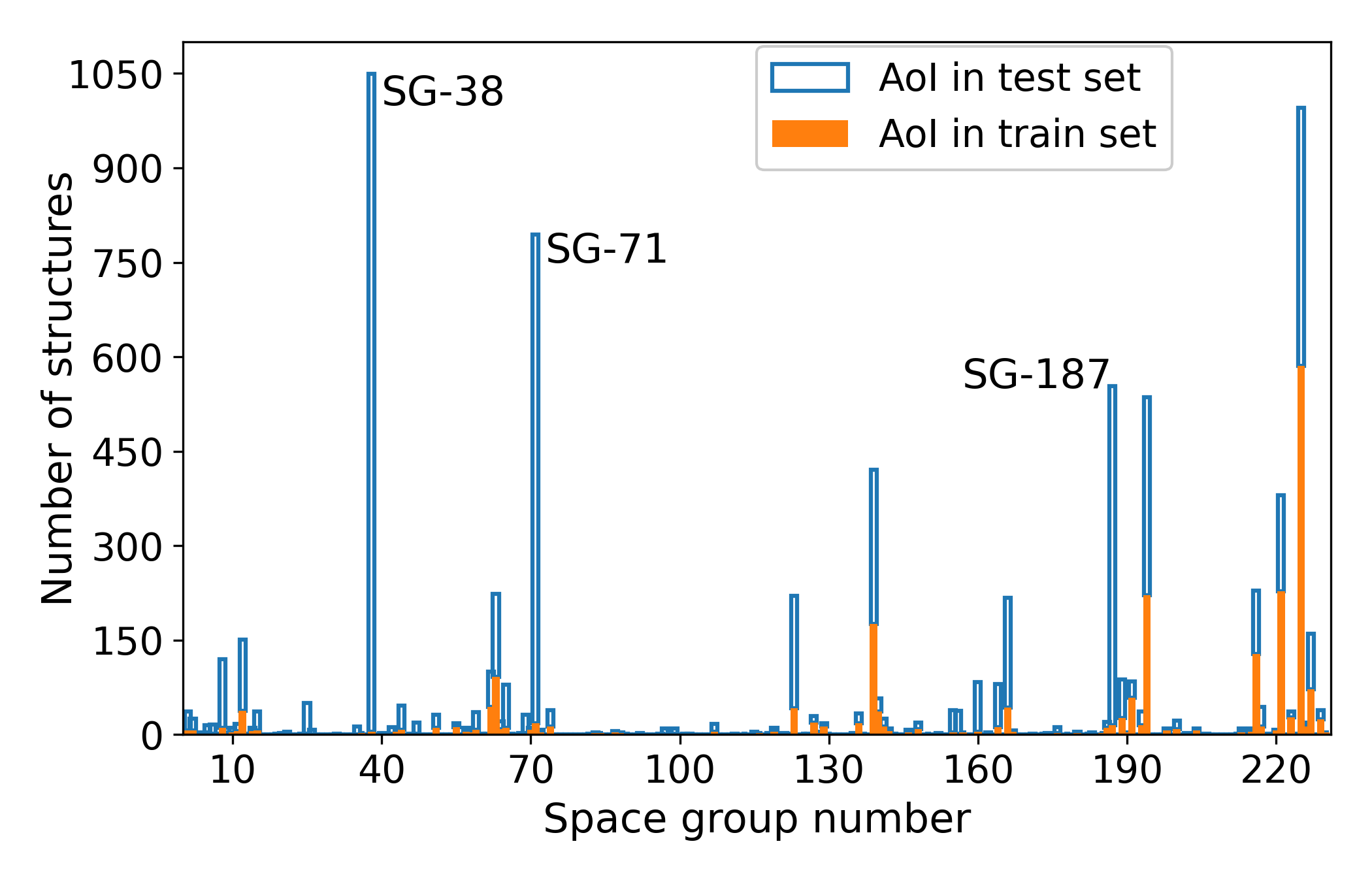}
	\caption{\label{fig:explain_issue_space_group} Number of structures as a function of space group number. For reference, the lattice type for a given interval of SG numbers is as follows: [1,2] triclinic, [3,15] monoclinic, [16,74] orthorhombic, [75,142] tetragonal, [143,167] trigonal, [168,194] hexagonal, [195,230] cubic. }
\end{figure}

\begin{figure*}
	\includegraphics[width=0.9\linewidth]{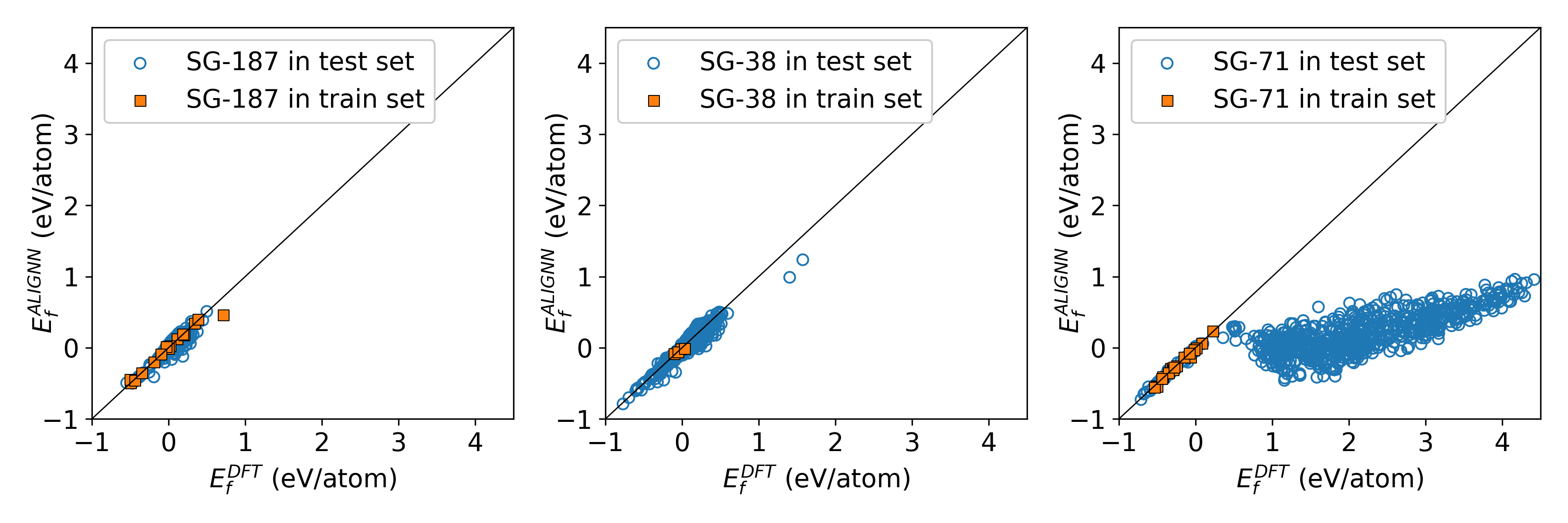}
\caption{\label{fig:explain_issue_space_group_parity_plot}Parity plot for the AoI data in different space groups.}
\end{figure*}

To study the data distribution in the structural space, we consider the crystallographic space group (SG) which describes the symmetry of a crystal. There are in total 230 SG for three-dimensional crystals, and the numbers of AoI belonging to these SG are shown in Fig.~\ref{fig:explain_issue_space_group}. It can be seen that there are few training data but much more test data for the SG-38, SG-71 and SG-187 structures. The parity plots for these structures are shown in Fig.~\ref{fig:explain_issue_space_group_parity_plot}. The formation energies for the 538 SG-187 structures in the test set are well predicted although there are only 15 training AoI with this SG. For the SG-38 AoI, the 1045 test samples that lie well beyond the small formation energy range of the 4 training data are also reasonably well predicted. By contrast, while the formation energies of the test SG-71 AoI in the $E_f^{\text{DFT}}$ range covered by the training data are well predicted, those with $E_f^{\text{DFT}}$ higher than 0.5 eV/atom are considerably underestimated by the ALIGNN-MP18 model. The different generalization behavior among these three SG suggests that failure to generalize is not strictly explained by underrepresentation of a given SG in the training data, nor by the range of target values.

\begin{figure}
	\includegraphics[width=\linewidth]{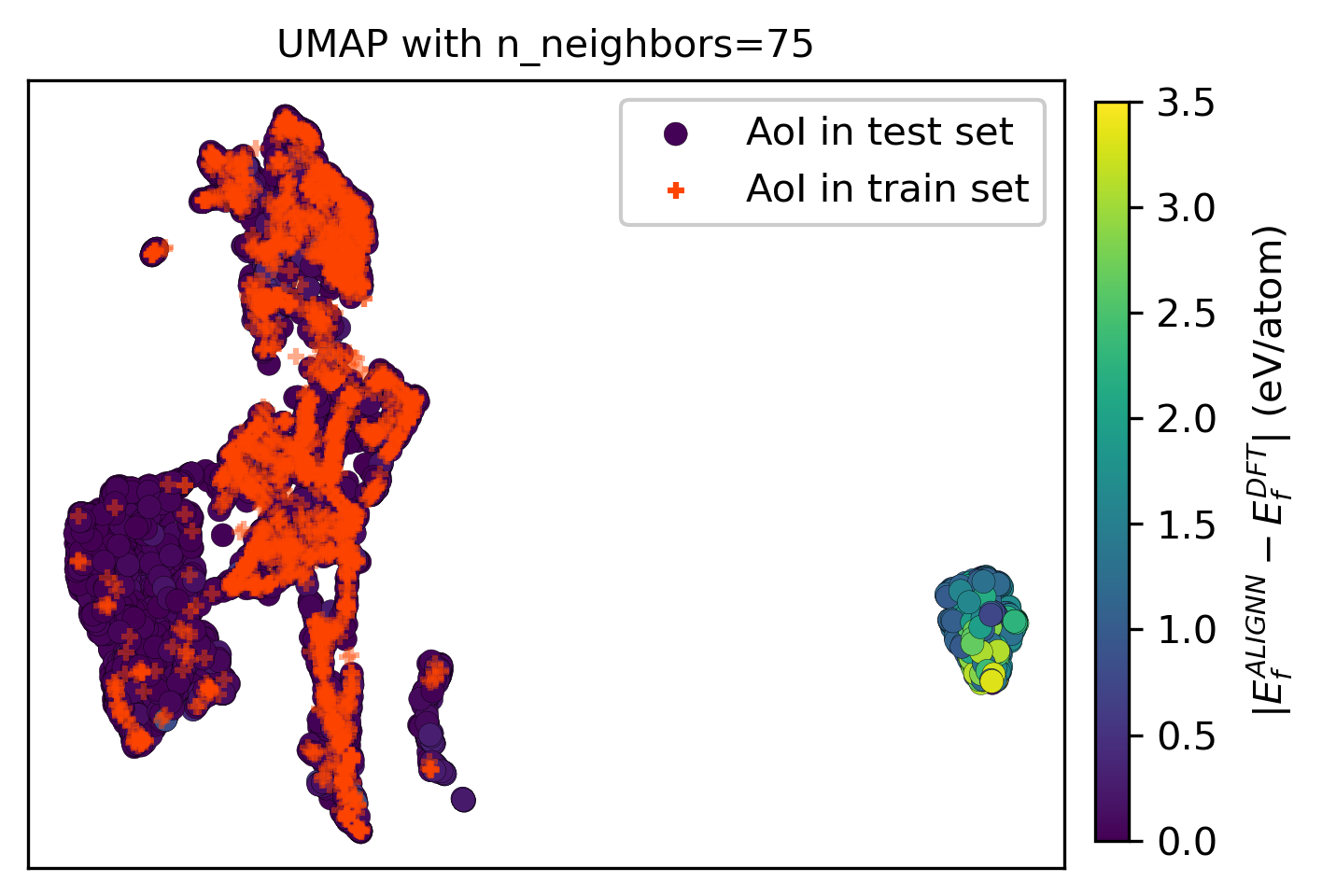}
\caption{\label{fig:explain_issue_UMAP} UMAP projection of the 90-dimensional feature space for the AoI training and test. The X and Y axis are not shown because the two dimensions in UMAP have no particular meanings. For the UMAP projection with only the test data, reader is referred to Supplemental Information Fig.~\ref{S-fig:SI_explain_issue_UMAP_no_train}.}
\end{figure}

While it is found that the poorly predicted data are primarily associated with ternary SG-71 structures, it is unclear why it is these structures that are particularly hard to predict for the ALIGNN model. It would be difficult to interrogate the ALIGNN model for a physical understanding of the problem. On the other hand, we find that there is a relatively strong correlation in the test set predictions between the ALIGNN and traditional ML models (Pearson's r for ALIGNN versus RF: 0.83, ALIGNN versus XGB: 0.77, ALIGNN versus LF: 0.68) and we can therefore use these models as surrogates for the ALIGNN to study the feature space in place of the neural network's representation. 

As mentioned in the previous section, there are 90 features after dropping the highly correlated ones from the initial set of 273 Matminer-extracted features. A typical way to understand high-dimensional data is to project them on a two-dimensional plane by applying dimension reduction. Here we use Uniform Manifold Approximation and Projection (UMAP), a stochastic and non-linear dimensionality reduction algorithm that preserves the data's local and global structure~\cite{umap}. One of the key hyperparameters in UMAP is \code{n\_neighbor} which constrains the size of the local neighborhood for learning the data's manifold structure. Lower values of \code{n\_neighbor} force UMAP to concentrate on the local structure of the data, whereas higher values push UMAP to provide a broader picture by neglecting finer details~\cite{umap}. By varying this hyperparameter, one can therefore obtain an idea of the data's structure at different scales. 
In Fig.~\ref{fig:explain_issue_UMAP}, we show a UMAP visualization of the feature space of the AoI training and test data. The test samples with low prediction errors are those clusters covered by the training data, whereas the majority of the poorly predicted alloys (which are largely SG-71 structures) form an isolated cluster away from the rest of the data. The Supplemental Information Fig.~\ref{S-fig:SI_explain_issue_UMAP_no_train} provides additional UMAP visualizations with smaller \code{n\_neighbor}, where there are smaller and more dispersed clusters.


It is worth noting that we have also attempted the commonly used principal component analysis (PCA) but found no clear clustering trend. This can be related to the fact that PCA is a linear algorithm and is not good at decoding the potentially non-linear relationships between features. Another reason may be that PCA looks for new dimensions that maximize the data's variance but does not preserve local topology of the data as UMAP does in Fig.~\ref{fig:explain_issue_UMAP}. 

\begin{figure*}[htbp]
	\includegraphics[width=0.57\linewidth]{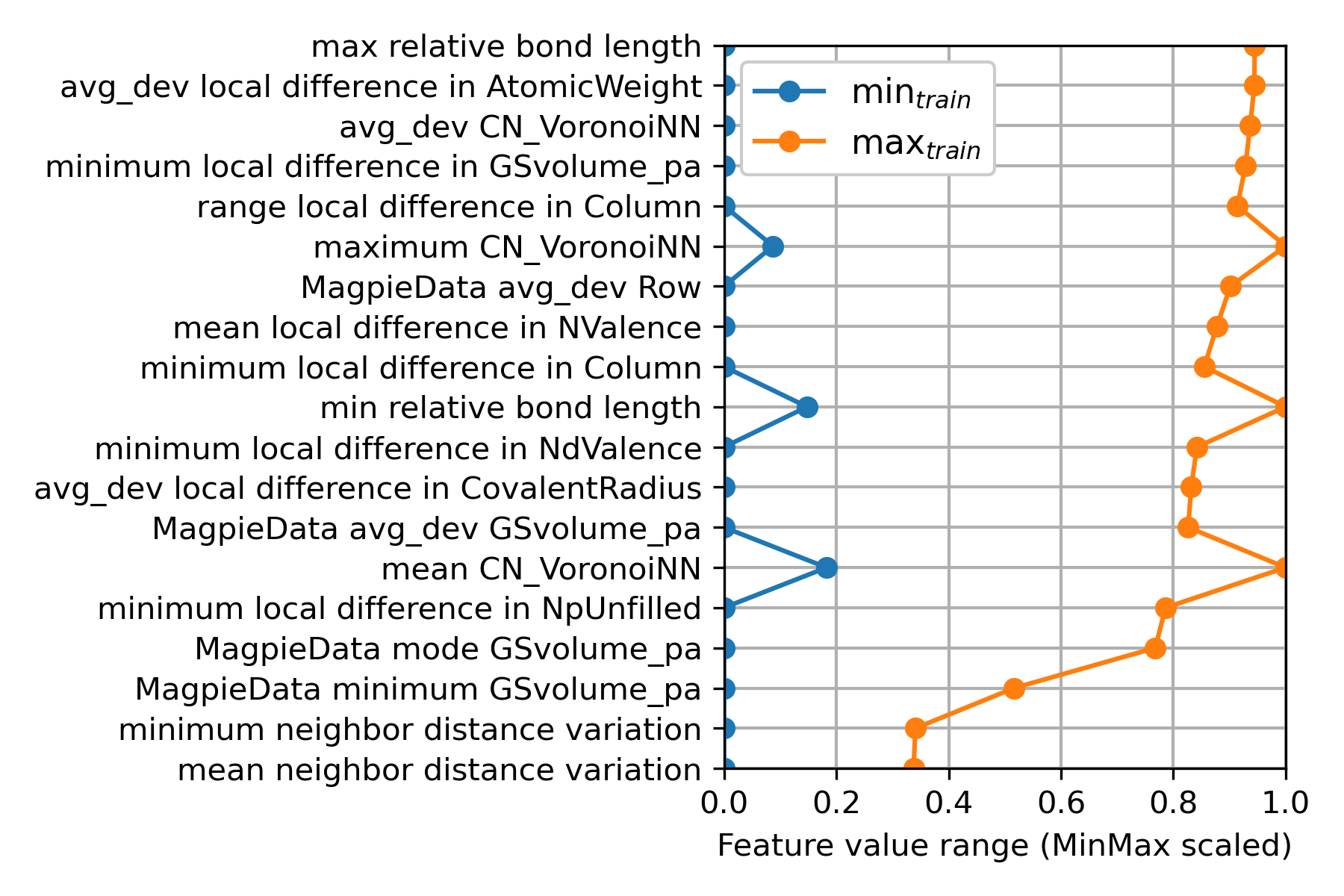}
	\includegraphics[width=0.42\linewidth]{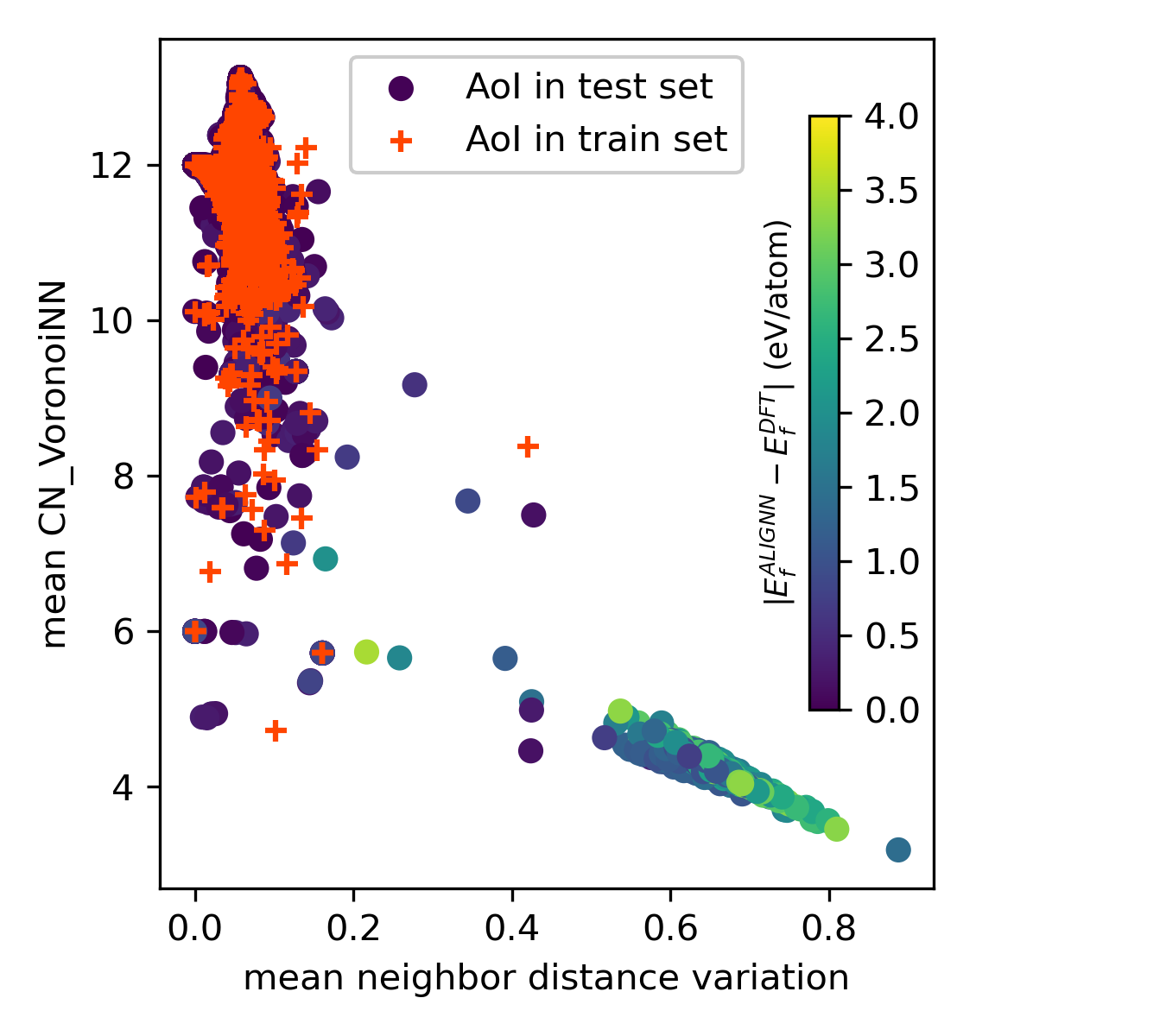}
	\caption{\label{fig:explain_issue_feature_range} Left panel: feature value range of the training AoI rescaled with respect to that of all the AoI data. Right panel: scatter plot of the AoI training and test data.}
\end{figure*}

Fig.~\ref{fig:explain_issue_UMAP} is a clear demonstration in the feature space that the poorly predicted test samples lie in an area well beyond that of the training AoI data. A complementary and more detailed understanding can be obtained by comparing the feature value ranges between the AoI training and test data. Fig.~\ref{fig:explain_issue_feature_range} shows the features whose ranges in the all AoI data are larger by more than 5~\% than the ranges in the training AoI data. Substantial changes in the value ranges can be noted for some features. 
In particular, only the lower 1/3 portion of \code{mean neighbor distance variation} and the higher 4/5 of \code{mean CN\_VoronoiNN} feature values are covered in the AoI training data. The feature \code{mean CN\_VoronoiNN} corresponds to the average number of nearest neighbors, while the feature
\code{mean neighbor distance variation} is the mean of the nearest neighbor distance variation, which measures the extent of atom displacement from high-symmetry sites and the extent of the lattice distortion against high-symmetry structures~\cite{Ward2017a,takahashi2020machine}. The right panel in Fig.~\ref{fig:explain_issue_feature_range} clearly reveals that the test data with large prediction errors have high \code{mean neighbor distance variation} and low \code{mean CN\_VoronoiNN} values, namely the poorly predicted structures are the ones with strong lattice distortion and a small number of nearest neighbors.


\subsection{Foreseeing performance issue}
\label{sec:spot_issue}
Our analysis in the previous section shows that ML models fail to generalize for compounds with large DFT formation energies relative to the range of formation energies in the training data. However, in a materials discovery setting we must foresee this generalization risk without such prior knowledge. In other words, it is important to identify the applicability domain and know whether ML models may be extrapolating and unreliable when used to explore unknown materials.

A natural idea is to define an applicability domain based on the training data density and coverage in the feature space, or equivalently estimate the similarity and distance between the training and test sets. However, this is not trivial in practice. While estimating data density based on basic compositional and structural features as shown in Fig.~\ref{fig:explain_issue_elements_dist} and~\ref{fig:explain_issue_space_group} could provide some indications of potential distribution shift, our discussion such as the one for Fig.~\ref{fig:explain_issue_space_group_parity_plot} also shows that fewer data for some SGs do not necessarily lead to poor predictions. Perhaps a more robust and comprehensive picture of the data can be obtained by extracting meaningful and predictive features and visualizing them with the aid of dimension reduction techniques such as UMAP. The distribution and clustering of the training and test data as shown in Fig.~\ref{fig:explain_issue_UMAP} can clearly help identify the test samples of which the ML predictions would be problematic. In addition, comparing the range of feature values in the training and test data (Fig.~\ref{fig:explain_issue_feature_range}) is a simple yet effective way to find out whether ML models are extrapolating when used to explore new regions of materials space. Various techniques including the above-mentioned ones should be used to inspect the training and the target space during the deployment of ML models, in order to reduce the risk of extrapolation in materials exploration.

\begin{figure}
	\includegraphics[width=\linewidth]{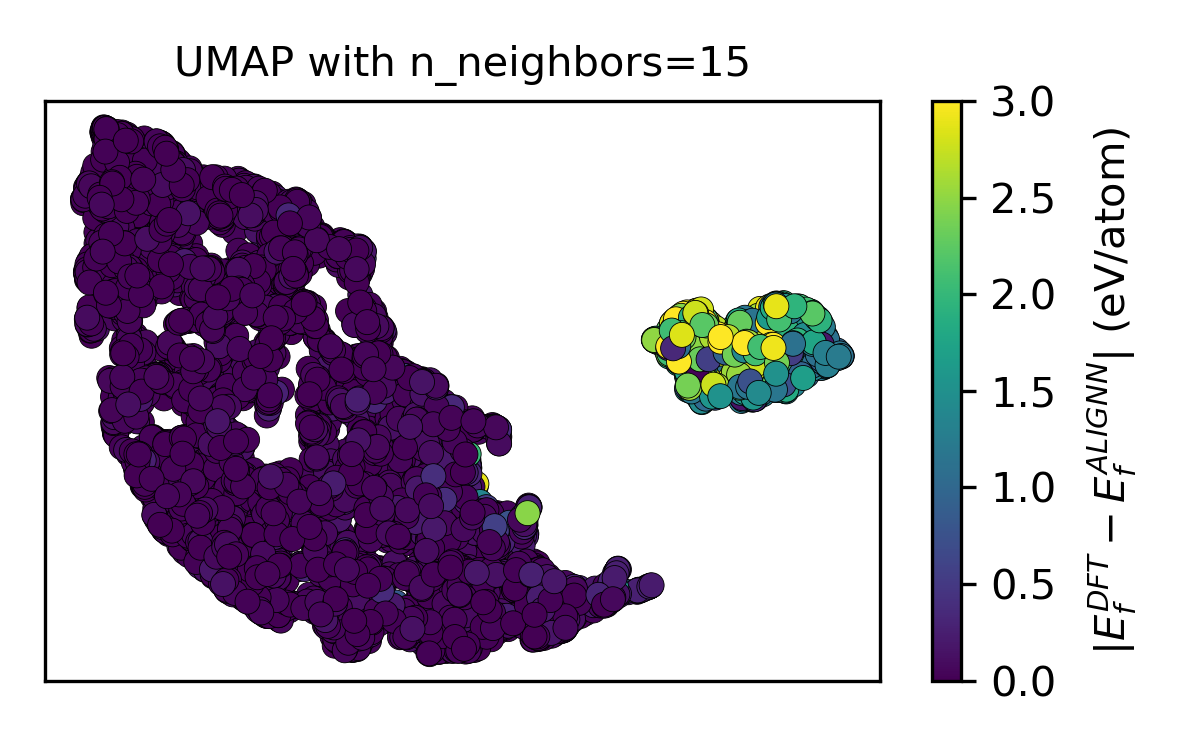}
	\caption{\label{fig:spot_issue_UMAP_error} UMAP projection of the AoI test data represented by the disagreement between different ML models. }
\end{figure}

Apart from carefully examining the feature space of datasets, one can also train multiple ML models and be more skeptical of the predictions of the test data with significant disagreement. For instance, our results in Fig.~\ref{fig:raise_issue_compare} show that different ML models show considerable disagreement for those out-of-distribution samples. Therefore, the degree of disagreements between the ML models can also be used to identify out-of-distribution samples. To better illustrate this point, we compute the prediction difference between ALIGNN and other models, namely $|E_f^{\text{ALIGNN}}$-$E_f^{\text{XGB}}|$, $|E_f^{\text{ALIGNN}}$-$E_f^{\text{RF}}|$, and $|E_f^{\text{ALIGNN}}$-$E_f^{\text{LF}}|$ for each of the test data. We then use UMAP to project the test data represented by the model disagreement in Fig.~\ref{fig:spot_issue_UMAP_error}, where the data are separated into two clusters. The cluster locating on the left is associated with test data having on average a much larger disagreement compared to the cluster on the right. Specifically, the mean value of $|E_f^{\text{ALIGNN}}$-$E_f^{\text{XGB}}|$ is 0.69 eV/atom (0.07 eV/atom) for the cluster locating on the left (right).

\begin{figure}
	\includegraphics[width=0.95\linewidth]{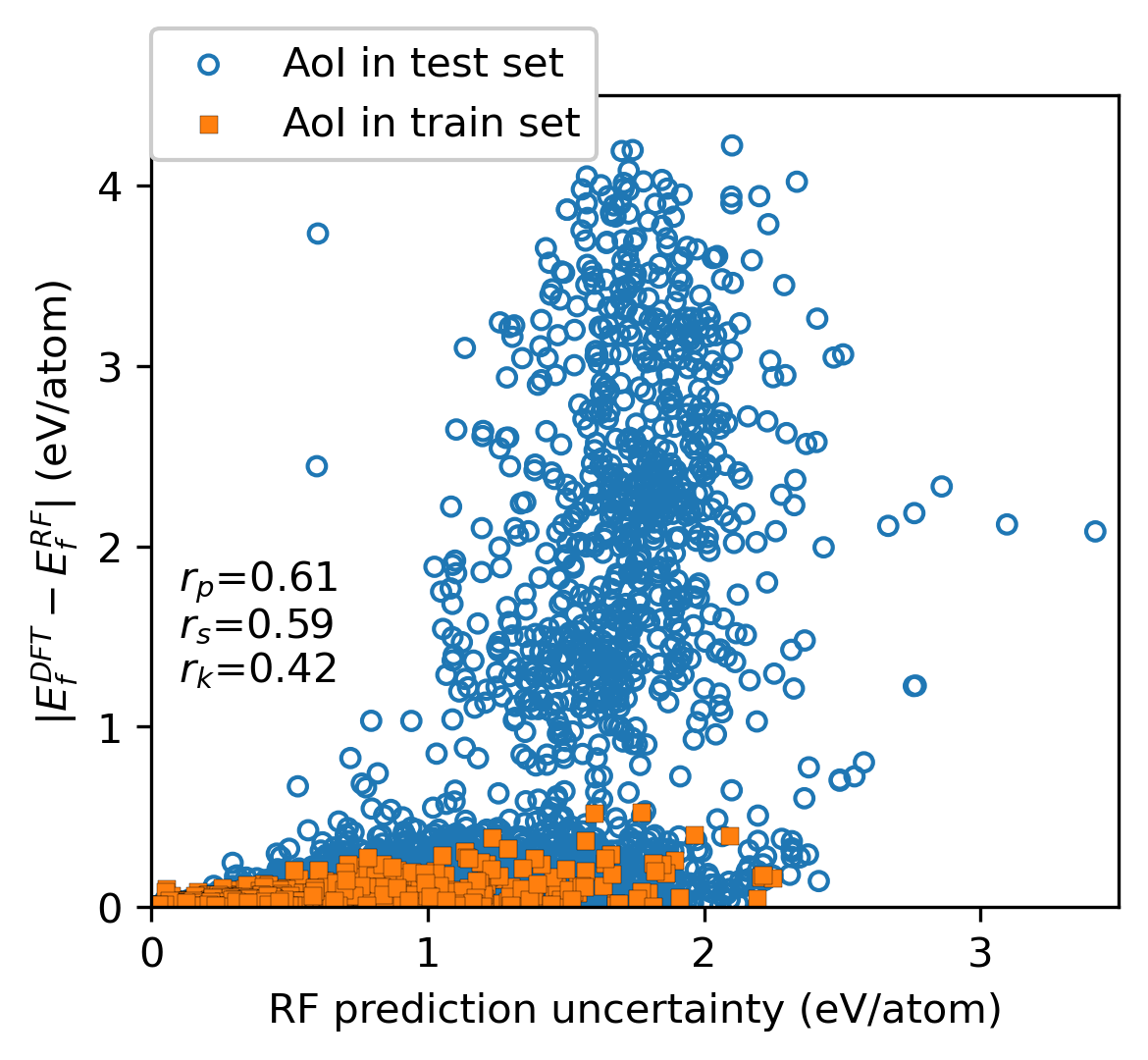}
	\caption{\label{fig:spot_issue_RF_uctt} Prediction uncertainty versus prediction error on the training and test AoI data for the MP18-pretrained RF model. The Pearson ($r_p$), Spearman ($r_s$), and Kendall ($r_k$) correlation coefficients for the prediction uncertainties and errors of the test AoI data are shown for reference. }
\end{figure}

Another commonly employed method to identify out-of-distribution test samples is to use uncertainty quantification. However, quantifying the uncertainty associated with the neural network predictions is challenging~\cite{Choudhary2022,abdar2021review} and is beyond the scope of this work. Instead, we consider the uncertainty associated with the RF model, based on the quantile regression forests~\cite{meinshausen2006quantile}. The prediction uncertainty of the RF model is computed as the width of the 95~\% confidence interval, namely the difference between the 2.5 and 97.5 percentiles of the trees' predictions. As shown in Fig.~\ref{fig:spot_issue_RF_uctt}, the RF uncertainty is only moderately correlated with the true prediction error for the test data. Based on the uncertainty distribution of the AoI in the training set, one may consider an uncertainty threshold between 1.5 eV/atom and 2.0 eV/atom for identifying samples that cannot be reliably predicted. However, using these thresholds not only includes many structures that actually have low prediction errors, but also excludes the poorly predicted structures whose prediction uncertainties are between 1.0 eV/atom to 1.5 eV/atom. Therefore, the RF uncertainty quantification does not allow to effectively discern the out-of-distribution from the in-distribution samples.

\subsection{Improving prediction robustness for materials exploration}
\label{sec:solve_issue}

Once we spot the gap between the training and test data, the next step is to improve prediction robustness by acquiring new data, ideally with a minimum of additional cost. In the following discussion of different acquisition policies, we consider the RF model as a proxy for the ALIGNN model, because it is much faster to update than the ALIGNN model and that its predictions have the best correlation with the ALIGNN predictions compared to the LF and XGB models. Active learning with the ALIGNN model is beyond the scope of this work.

\begin{figure}[btp]
	\includegraphics[width=\linewidth]{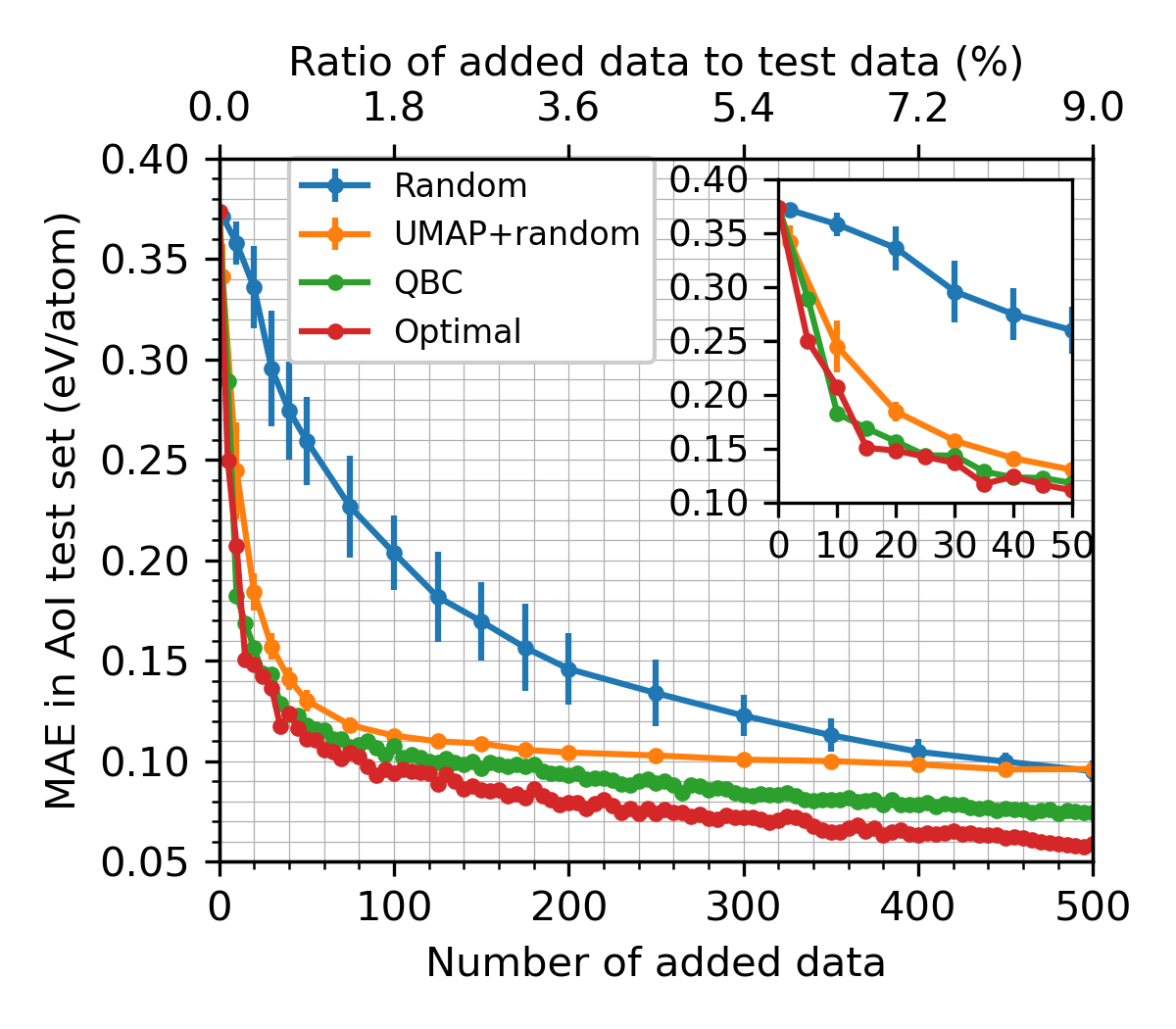}
	\caption{\label{fig:solve_issue_RF_AL0} Test MAE as a function of number of selected test data added to the training set. The inset shows the enlarged region at the early stage of the active learning process. All the results are reported for the RF model.}
\end{figure}

Our discussion in the previous sections can provide insights for establishing the acquisition policy. For instance, one can prioritize the UMAP space poorly covered by the training data. In Fig.~\ref{fig:solve_issue_RF_AL0} we demonstrate the effectiveness of this simple idea. We add a given number of samples randomly taken from the isolated cluster in the UMAP plot (Fig.~\ref{fig:explain_issue_UMAP}) to the original MP18 training set to train the RF model. We find a significant decrease in the test MAE, compared with the baseline acquisition policy of randomly taking data from the whole test set. With only 50 data (out of 5539 test data) added, the UMAP-guided random sampling leads to a test MAE of 0.13 eV/atom, which is only half of the test MAE of 0.27 eV/atom resulting from the baseline policy (random sampling) with the same number of added samples. The latter needs five times the number of samples to arrive at the same MAE. 

As discussed in Fig.~\ref{fig:spot_issue_UMAP_error}, the level of disagreement between the ML models is also useful in finding the poorly predicted samples. We therefore consider the query by committee (QBC) acquisition, where we select the test data that have the strongest disagreement among the three committee members (RF, LF and XGB). As shown in Fig.~\ref{fig:solve_issue_RF_AL0}, the QBC strategy shows a slightly better performance than the UMAP-guided random sampling. Hoping to find an even better performance in the early acquisition stage, we further consider combining the QBC with the UMAP-guided sampling, but find the resulting performance is similar to using only the QBC strategy. To estimate whether this is because we are reaching the optimal strategy, we compute another acquisition curve, where we select samples that have the largest RF-DFT disagreement. As the DFT labels are assumed known, this curve is not regarded as a true active learning acquisition, but is only used to estimate the optimal performance that an active learning can reach. It is clear from Fig.~\ref{fig:solve_issue_RF_AL0} that the QBC curve is quite close to the estimated optimal curve, so it is not surprising that combining it with UMAP does not bring further improvement.

It is worth noting that with the UMAP-guided sampling or the QBC policy, adding only 1~\% of the data already results in a reasonable test MAE. Therefore, these two strategies are very effective in identifying the most diversified and informative samples. Though Fig.~\ref{fig:solve_issue_RF_AL0} shows that adding even more samples can further improve the model performance in the AoI subspace, such an improvement is rather incremental. The compute should be saved to explore the regions of materials space that could bring potentially drastic gain in the prediction robustness and accuracy.

In case acquiring new data is not possible, prediction robustness for those out-of-distribution samples can still be improved by using more extrapolative models. For instance, tree-based models are usually considered to be interpolative, as is also found here for the RF model (Fig.~\ref{fig:raise_issue_compare}). By simply adding a linear component to the RF model, however, our LF model gives a more robust estimation of the stability for those out-of-distribution samples. The better extrapolation performance is enabled by the features whose ranges in the test set far exceed those in the training set, as removing these features from the LF model reduces the extrapolation performance to the same level of the RF model. On the other hand, the better extrapolation performance of the XGB and LF models also comes from the training data outside the space of interests (namely non-AoI), since training only on the MP18 AoI data leads to performance similar to that of the RF model. This indicates that ML models can learn from less relevant data outsides the target space for better generalization performance.

\subsection{Conclusion}
This work is focused on the prediction robustness of ML models, by examining the formation energy predictions of the MP18-pretrained models for the new alloys in the latest MP21 dataset. We considered the ALIGNN model, a graph neural network with a state-of-the-art performance in the Matbench formation energy prediction task, as well as three traditional descriptor-based ML models (XGB, RF and LF). Despite the excellent test performance in the MP18, the MP18-pretrained ALIGNN model strongly underestimated the DFT formation energies of some test data in the MP21. While this performance issue was also found in the traditional ML models, the XGB and LF models provided more robust phase stability estimation for the test data. We analyzed and discussed the origins of the performance degradation from multiple perspectives. In particular, we used UMAP to perform dimension reduction on the high-dimensional Matminer-extracted features, revealing that the poorly predicted data lie far beyond the feature space occupied by the training set. With these insights, we then discussed possible methods, including UMAP-aided clustering and cross-checking multiple ML models, to identify out-of-distribution data and foresee performance degradation. Finally, we provided suggestions to improve prediction robustness for materials exploration. We showed that the accuracy can be greatly improving by just adding a very small amount of new data as identified by UMAP clustering and querying different ML models. We believe that UMAP-guided active learning shows promising potential for future dataset expansion. In cases where data acquisition is not possible, we also propose to include extrapolative components such as linear models for a more robust prediction for out-of-distribution samples. We hope this work can raise the awareness of the limitations of the current ML approaches in the materials science community, and provide valuable insights for building databases and ML models with better prediction robustness and generalizability.

\section{Methods}

The 2018.06.01 snapshot of Materials Project is retrieved by using JARVIS-tools~\cite{Choudhary2020}, while the latest 2021.11.10 version is retrieved by using the Materials Project API~\cite{Jain2013}. For each material, Materials Project uses the \code{material\_id} field as its identifier and the \code{task\_ids} field to store its past and current identifiers. The structures in the MP21 \code{task\_id} field contains an MP18 identifier are considered as the materials existing in the MP18, whereas the rest in the MP21 are considered to be the new materials unseen in the MP18.

We use the ALIGNN-MP18 model that was published with the original paper~\cite{Choudhary2021}. We use Matminer~\cite{Ward2018} to extract 273 compositional and structural features~\cite{Ward2017a}, and obtain 90 features after sequentially dropping highly correlated features (with a Pearson's r of 0.7 as the threshold). We use three traditional ML models: the gradient-boosted trees as implemented in \code{XGBoost} (XGB)~\cite{xgboost}, random forests (RF) as implemented in \code{scikit-learn}~\cite{sklearn}, and linear forests (LF) as implemented in \code{linear-forest}~\cite{linearforest,Zhang2017}. For the XGB model, we use 2000 estimators, a learning rate of 0.1, an L1 (L2) regularization strength of 0.1 (0.0), and the histogram tree grow method. For the RF model, we use 100 estimators, 30~\% of the features for the best splitting. We combine the same RF model with a Ridge model with a regularization strength of 1 for the LF model. We use the packages' default settings for other hyperparameters not mentioned here. 

\section*{Data availability}
The data required and generated by our code are publicly available at url (to be inserted upon acceptance of the paper).

\section*{Code availability}
The code for ML training, analysis, and figure generations in this work is available on GitHub at url (to be inserted upon acceptance of the paper).

\begin{acknowledgments}
We acknowledge funding provided by Natural Resources Canada’s Office of Energy Research and Development (OERD).  \copyright His Majesty the King in Right of Canada, as represented by the Minister of Natural Resources, 2022.
\end{acknowledgments}

\section*{Author contributions}
K.L., B.D. and J.H.S. conceived and designed the project. K.L. trained ML models, analyzed results, and drafted the manuscript. J.H.S. supervised the project. K.L., B.D., K.C. and J.H.S. discussed the results. B.D., K.C., M.G. and J.H.S. reviewed and edited the manuscript. All authors contributed to the manuscript preparation.

\section*{Competing interests}
The authors declare no competing interests.

\bibliography{lib} 
\end{document}